\documentclass[prl,twocolumn,showpacs]{revtex4}

\usepackage{graphicx}
\usepackage{bm}
\begin{document}
\title{Full Current Statistics of Incoherent "Cold Electrons"}
\author{Dmitry A. Bagrets}
\affiliation{Institut f\"ur Theoretische Festk\"orperphysik, 
Universit\"at Karlsruhe, 76128 Karlsruhe, Germany.
}
\date{\today}
\pacs{73.23.-b, 72.70.+m, 05.40.-a, 73.50.Td}

\begin{abstract}
We evaluate the full current statistics (FCS) in the low dimensional ( 1D and 2D ) diffusive
conductors in the incoherent regime, $eV\!\gg\! E_{\rm Th}=D/L^2$, $E_{\rm Th}$ being the
Thouless energy. It is shown that
Coulomb interaction substantially enhances the probability of big current fluctuations for short 
conductors with $E_{\rm Th}\!\gg\!1/\tau_E$, $\tau_E$ being the energy relaxation time,
leading to the exponential tails in the current distribution.
The current fluctuations are most strong for low temperatures, 
provided
$E_{\rm Th}\sim [(eV)^2/D\nu_1^2\bigr]^{1/3}$ for 1D and 
$E_{\rm Th}\sim  (eV/g)\ln g$ for 2D, where $g$ is a dimensionless conductance 
and $\nu_1$ is a 1D density of states. The FCS in the "hot electron" regime
is also discussed.

\end{abstract}

\maketitle

The influence of the Coulomb interaction onto the transport properties of low-dimensional
diffusive systems has been a subject of extensive research for more than twenty 
years~\cite{El_El, Lee}. Initially, the conductance only was a main object of study.
The powerful alternative to that is to investigate a quantum noise~\cite{BlanterReview, QNoise},
or, more generally, the full current statistics (FCS)~\cite{Levitov}.

  In the short diffusive wires with $E_{\rm Th}\gg eV$,
where $E_{\rm Th}=D/L^2$ is the Thouless energy, 
%$D$ is a diffusion constant and $L$ is a size of the conductor,  
the shot noise equals to
$S= 2|e|I F$, $F=1/3$ being the Fano factor~\cite{Beenakker, Devoret}. In this case
the conductor is coherent and effectively zero-dimensional so that
all effects of Coulomb interaction come from
the external electromagnetic environment. It modifies 
the conductance, noise~\cite{Zaikin} and generally the FCS~\cite{Kin_Naz, Bag_Naz}.

Much less is known about the role of 
Coulomb interaction onto the FCS in the quasi- 1D and 2D diffusive systems, when
$E_{\rm Th}\ll eV$. Under this condition the inelastic electron-electron scattering
inside the conductor is important. This subject has recently attracted the attention
in Ref.~\cite{Gefen, Pilgram, Mirlin}, where the "hot electron" regime in the FCS was discussed.
In this regime $E_{\rm Th}\ll 1/\tau_E$, $\tau_E$ being the energy relaxation
time, and the electron distribution function relaxes to the
local Fermi distribution. This changes the Fano factor $F$ from $1/3$ to 
$\sqrt{3}/4$~\cite{Hot}, that was confirmed experimentally~\cite{Devoret}.

 The microscopic theory~\cite{AA} of electron-electron interaction  in low-dimensional
disordered conductors predicts, however, {\it two} different time scales,
$\tau_\phi$ and $\tau_E$, where $\tau_\phi \ll \tau_E$ is a dephasing time (See Table I). 
It is usually believed~\cite{Aleiner} that classical phenomena described by the 
Boltzmann equation are governed by $\tau_E$.
While the time $\tau_\phi$ manifests itself in essentially quantum-mechanical
phenomena.
Since the FCS is a classical quantity one might expect it to cross over 
between the coherent and the "hot electron" regime
on the scale $E_{\rm Th}\sim 1/\tau_E$.

\begin{table}[b]
\caption{\label{tab:table1}
The electron scattering times for low-dimensional (1D and 2D) diffusive 
conductors, $E=\max\{T,eV\}$.
At $T\lesssim\tau^{-1}_\phi(V)$, we get $\tau^*=\tau_\phi(V)$.
}
\begin{ruledtabular}
\begin{tabular}{lccc}
d & $1/\tau_E$ & $1/\tau_\phi$ & $1/\tau^*,\,\, T\gtrsim \tau^{-1}_\phi(V)$ \\
\hline
1 & $(E/D)^{1/2}\nu_1^{-1}$ &  $(E^2/D\nu_1^2)^{1/3}$ & $(eV/T)^{1/2}\,\,\tau_E^{-1}(V) $ \\
2 & $E/g$ & $(E/g)\ln g$ & $\ln(eV/T) \,\, \tau_E^{-1}(V)$ \\
\end{tabular}
\end{ruledtabular}
\end{table}

In this paper we consider the FCS in the low dimensional ( $d=1,\,2$ ) diffusive
conductors, taking into account the Coulomb interaction.
We show that the time $\tau_E$ is indeed responsible for the smooth
crossover between the coherent and the "hot electron" limits if one considers 
the noise and the 3$^d$ cumulant. However it is not the case for the higher order
cumulants of charge transfer in the shot noise limit $eV\gg T$.  
Moreover, in this limit the smooth crossover in the FCS does not exist.
The Coulomb interaction drastically enhances the probability of 
current fluctuations for short conductors $E_{\rm Th} \gg 1/\tau_E$.
In this regime the higher order cumulants 
are given by
\begin{equation}
\langle\!\langle n^{2k, 2k+1} \rangle\!\rangle \propto
\frac{\langle n\rangle}{g} 
\Bigl(\frac{\displaystyle eV}{\displaystyle E_{\rm Th}}\Bigr)^{d/2}
\Bigl(\frac{\displaystyle eV}{\displaystyle \omega^*}\Bigr)^{k-1-d/2}, \,\,k>2
\label{cumulants}
\end{equation}
where $\langle n\rangle \gg 1$ is the average number
of electrons transfered,
$g\gg 1$ is a dimensionless conductance, 
$\,\omega^* = \max\{E_{\rm Th}, T,\epsilon^*\}$ and the scale $\epsilon^*$ reads
\begin{equation}
\epsilon^*(V) \simeq \left\{
\begin{array}{cc}
(eV)^2/ g^2\, E_{\rm Th}, &  \,\,1D \\
 eV \exp\{ - g\,E_{\rm Th}/eV \} , & \,\, 2D.
\end{array} \right.
\label{E_star} 
\end{equation}
Eq.(\ref{cumulants}) shows that
each $(k+1)$-th cumulant of charge transfer is parametrically enhanced versus the $k$-th one
by the large factor $eV/\omega^*\gg 1$. It also follows from Eq.(\ref{cumulants}) that
the higher cumulants grow with increasing the voltage at
$E_{\rm Th}>1/\tau^*$ and decay at $E_{\rm Th}<1/\tau^*$, where
the new time scale $\tau^*(eV,T)$ is parametrically smaller than $\tau_E$, $\tau^*\ll \tau_E$.
(See Table I). 
The current fluctuations are most strong, provided
$T\lesssim E_{\rm Th} \sim 1/\tau_\phi(V)$. 
Therefore at the strongly non-equilibrium
situation the time $\tau_\phi$ rather than $\tau_E$ governs the crossover in the FCS
between the coherent and the "hot electron" limits.

{\it Model and the effective action.}
We consider a quasi- one-dimensional (1D) diffusive
wire of a length $L$ and a quasi- two-dimensional (2D) film of a size $L\times L$, 
with dimensionless conductance $g\gg 1$ and diffusion coefficient $D$.
They are attached to two reservoirs with negligible external impedance 
which are kept at voltages $\pm V/2$.
The current flows along $z$ direction.
We assume the incoherent regime, $\max\{eV,T\}\gg E_{\rm Th}$, and 
disregard the possible electron-phonon scattering, so that $L\ll L_{\rm e-ph}$.

 Our goal is to evaluate the cumulant generating function (CGF) $S(\chi)$.
The Fourier transform of $\exp(-S)$ 
with respect to the "counting filed" $\chi$
gives the current probability distribution $P(I)$ (See~\cite{Levitov}). 
The derivatives of $S$ give the average value of current,
shot-noise and higher order moments 
$\langle\!\langle n^k \rangle\!\rangle$ of charge transfer
during the observation time $t_0$.

  To evaluate the CGF, taking into account the Coulomb interaction, we have used the
Keldysh technique and employed
the low-energy field theory of the diffusive transport~\cite{Kamenev} with the action
\begin{eqnarray}
F[\chi, Q, {\bf A}] = \frac{1}{8} g L^{2-d} \int d^{\,d}{\bf r}\, {\rm Tr} 
\left( {\bf\nabla}Q -i[\hat A , Q] \right)^2 + \label{Action} \\
 2 i\pi\nu_d \int d^{\,d}{\bf r}\, {\rm Tr}\Bigl( i\partial_t Q\Bigr)
- \frac{i}{8\pi e^2} \int\limits_{-\infty}^{+\infty} d\,t \int d^3{\bf r} 
\Bigr( \dot{\bf A}_1^2  - \dot{\bf A}_2^2 \Bigl)
\nonumber
\end{eqnarray} 
Here  $\hat A = {\rm diag}( {\bf A}_1(t, {\bf r}), {\bf A}_2(t, {\bf r}))$ is the 
$2\times 2$ matrix in Keldysh space, where ${\bf A}_{1,2}$ stand for 
fluctuating vector potentials in the conductor. Hereafter we use 
a longitudinal photon field,  
${\rm curl}\,{\bf A}=0$, thus neglecting the relativistic effects.

The matrix $\hat Q({\bf r}, t_1, t_2)$ accounts for the electron degrees of freedom
and obeys  the semi-classical constrain $\hat Q({\bf r}) \circ \hat Q({\bf r}) = \delta(t_1-t_2)$. 
The action $F$ depends on $\chi$ via the boundary conditions
imposed on the field $Q$ at the boundaries with the left(L) and right(R) reservoirs~\cite{Belzig}:
$Q\bigr|_{{\bf r}=R} = \hat G_R$ and 
$Q\bigr|_{{\bf r}=L} = \hat G_L(\chi) =
\exp (i \chi \hat \tau_3/2) \hat G_L \exp (-i \chi \hat \tau_3/2)$.
Here $G_{L,R}$ are the Keldysh Green functions in the leads. 

  With action (\ref{Action}) the CGF should be evaluated as a path integral
over all possible realization ${\bf A}_{1,2}$ and $\hat Q$. 
We proceed along the lines of Ref.\cite{Kamenev} and employ the parameterization
$Q = e^{i W} \hat G e^{-i W}$,  $W\hat G + \hat G  W=0$. Here field $W$ 
accounts for the rapid fluctuations of $Q$ with typical frequencies $\omega \gg E_{\rm Th}$ 
and momenta $q\gg 1/L$, while $\hat G(\epsilon,{\bf r})$ is 
the stationary Green  function 
varying in space on the scale $\sim L$.
First we integrate out the field $W$ in the Gaussian
approximation to obtain the non-linear action ${\widetilde F}(\chi,\hat G, {\bf A})$ of 
the screened electromagnetic fluctuations. 
We keep only quadratic terms to ${\widetilde F}$ that is equivalent
to the random phase approximation (RPA). At the second step one can integrate the photon
field $\bf A$ and arrive to the effective action 
$F_{\rm eff}[\chi, \hat G]$.
Then the saddle point approximation,
$\delta F_{\rm eff}[\chi, \hat G]/\delta \hat G = 0$  yields the  kinetic equation
for $\hat G(\epsilon, {\bf r})$.

   For the rest we restrict consideration to the universal limit of a short screening radius
$r^{-1}=(4\pi e^2\nu_3)^{1/2}\gg\sqrt{eV/D}$. In this limit we get the answer
\begin{eqnarray}
F_{\rm eff}[\chi, \hat G] = \frac{t_0}{8} g L^{2-d} \int d^{\,d}{\bf r}
\int\frac{d\,\epsilon}{2\pi}\, {\rm Tr} 
\left( {\bf\nabla} \hat G_\epsilon({\bf r})  \right)^2 +  \label{Eff_Action} \\
+\frac{t_0}{2}\int d^{\,d}{\bf r}\int \frac{d\omega\,d^{\,d}\bf q}{{(2\pi)}^{d+1}}
{\rm ln}\,\left[
\frac{{\rm Det}\,\bigl|\bigl| {\cal D}^{-1}_\omega({\bf r, q})\bigr|\bigr|}
{-\bigl((D{\bf q}^2)^2+\omega^2\bigr)} 
\right]
\nonumber
\end{eqnarray} 
where ${\cal D}_\omega$ is $2\times 2$ matrix operator in Keldysh space, 
corresponding to the non-equilibrium diffuson propagator:
\begin{eqnarray}
&&{\cal D}^{\alpha\,\beta}_\omega({\bf r, q}) 
= \Bigl[ D{\bf q}^2 \,\tau_1^{\alpha\,\beta} + \label{Diff}\\
&&\frac{i}{4}\int d\epsilon\, {\rm Tr}\left( 
\gamma^\alpha \gamma^\beta - 
\gamma^\alpha \hat G_{\epsilon+\omega/2}({\bf r}) 
\gamma^\beta \hat G_{\epsilon-\omega/2}({\bf r})
\right)
 \Bigr]^{-1}  \nonumber
\end{eqnarray}
%with $\alpha,\beta = 0,1$ and 
with $\gamma^0=\hat 1$, $\gamma^1=\hat\tau_3$.
To derive the action~(\ref{Eff_Action}) we have used a local approximation, i.e. we
neglected gradient corrections proportional to
$(\nabla\hat G\sim1/L) \ll \nabla W$. 

\begin{figure}[t]
\includegraphics[width=3.4in]{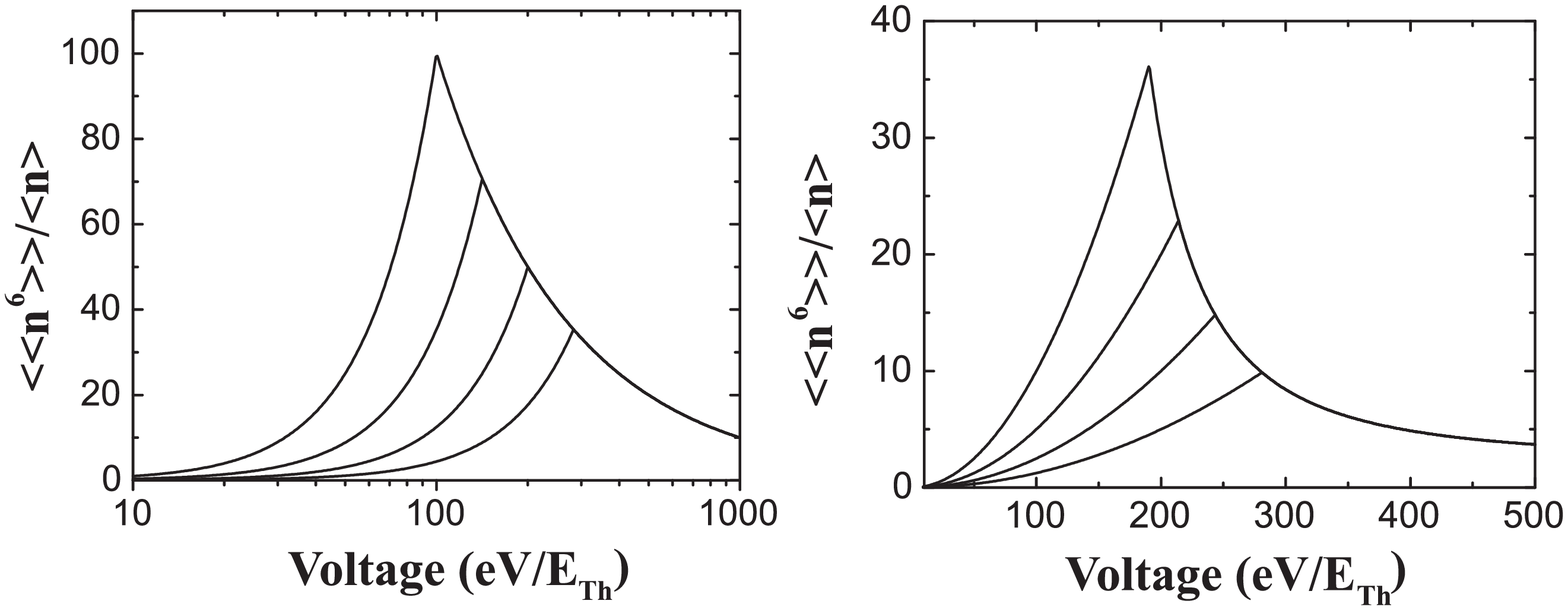}
\caption{ The sketch of voltage dependence of the 6$^{th}$ cumulant of charge transfer.
Left plot - 1D, conductance $g=100$, right plot - 2D, $g=10^3$.
The temperature changes from up to down, $T/E_{\rm Th} = 1,\, 2,\, 4,\, 8$.
} 
\end{figure}

Minimizing the action $F_{\rm eff}$ under constrain
$\hat G(\epsilon, {\bf r})^2 = 1$ one can obtain the 
non-linear kinetic equation for $G(\epsilon,{\bf r})$ in the
form 
$D\,\nabla\left(\hat G_\epsilon({\bf r})\nabla \hat G_\epsilon({\bf r})\right) = 
\left[\hat{\cal I}_\epsilon({\bf r}), \hat G_\epsilon({\bf r})\right]
$, where
\begin{eqnarray}
\hat{\cal I}_\epsilon({\bf r}) &=&  \frac{i}{8\nu_d}
\sum_{\alpha,\beta} 
\int \frac{d\omega\,d^{\,d}\bf q}{{(2\pi)}^{d+1}}
{\cal D}^{\alpha\,\beta}_\omega({\bf r}, {\bf q}) \label{St} \\
&\times& 
\left(\gamma^\alpha \hat G_{\epsilon-\omega}({\bf r})\gamma^\beta 
+\gamma^\beta \hat G_{\epsilon+\omega}({\bf r})\gamma^\alpha 
\right) \nonumber
\end{eqnarray}
is the matrix collision integral.
This kinetic equation should be supplemented by the $\chi$-dependent boundary conditions
at the interfaces with the leads.
The action~(\ref{Eff_Action}, \ref{Diff}) is one of 
the central results of the paper. 
%The CGF can be evaluated from it using 
%the solution $G_\epsilon(\chi, {\bf r})$ of kinetic equation.
In the absence of "counting" field our matrix kinetic equation
reduces to the standard kinetic equation for the distribution function with a singular kernel 
$K(\omega) \propto \omega^{\,d/2-2}$ (See~\cite{AA, Aleiner, Kamenev}).
We also note that only  real inelastic processes with energy transfer
$\omega \lesssim \max\{eV,T\}$ contribute to the action~(\ref{Eff_Action}).
In the rest of the paper we consider the shot-noise limit $eV\gg T$ only
and proceed with

{\it Incoherent "cold electron" regime,} $E_{\rm Th} \gg 1/\tau_E$. 
In this regime the collision term in kinetic equation is small and 
one can try  to find the Green function 
perturbatively around the coherent solution 
which obeys the Usadel equation 
$\nabla_z\left(\hat G^{\,0}_\epsilon(z)\nabla_z \hat G^{\,0}_\epsilon(z)\right)=0$.
%Here $0<z<1$ is a dimensionless coordinate along the current direction. 
In the first order in $1/(E_{\rm Th}\tau_E)$ the CGF 
can be found  by substituting $\hat G^0$ in the
action~(\ref{Eff_Action}). The main contribution 
comes from frequencies $T<\omega<eV$.
After some algebra we obtain
\begin{equation}
S(\chi)= -\frac{t_0}{8\pi} eV g\,\,
\theta^2_\chi(\epsilon) +F_{\rm Coll}(\chi).
\label{S_chi}
\end{equation}
Here
$\theta_\chi=\ln(u+\sqrt{u^2-1})$, $u=2e^{i\chi}-1$
and
\begin{equation}
F_{\rm Coll} = \frac{t_0 L^{d}}{2}
\int_0^1 \! dz\!\!\!\!\!\!\!\int\limits_{\omega^*\lesssim\omega < eV}\!\!\!\!\!
\frac{ d\omega\,d^{\,d} {\bf q} }{ (2\pi)^{d+1} }
{\rm ln}\,\biggl\{ 
1 - \frac{N_\omega\, \omega^2 \Pi(\chi,z)}{(D{\bf q}^2)^2+\omega^2} 
\biggr\} \label{F_Coll} 
\end{equation}
\begin{eqnarray}
\Pi(\chi, z) &=&  - 4 \, L_\chi(z) R_\chi(z) e^{i\chi}
\Bigl\{ 1\!-\!L_\chi(z)\!\!-\!\!R_\chi(z) -  \nonumber \\
&& [ z L_\chi(z) \!+\!(1-z)R_\chi(z)\Bigl](e^{i\chi}\!-\!1)\Big\}
\label{Pi}
\end{eqnarray}
where $N_\omega = (eV/|\omega|-1)$, 
$L_\chi(z)=\sinh(1-z)\,\theta_\chi /\sinh \theta_\chi$,
$R_\chi(z)=\sinh z\,\theta_\chi /\sinh \theta_\chi$ and
$\omega^* = \max\{E_{\rm Th}, T\}$. The presence of Thouless energy in 
the low frequency cut-off $\omega^*$ is due to the fact that the lowest allowed momenta $q$
in the diffusion propagator equals to $1/L$, while $T$ takes into account the smearing
of a step in the Fermi distribution.

The result~(\ref{F_Coll}) with the above defined cut-off $\omega^*$ ceased to be valid 
at sufficiently high voltages. 
Indeed, substituting
a zero order distribution function 
$f_0(\epsilon) = (1-z) f_F(\epsilon-eV/2) + z f_F(\epsilon+eV/2)$ 
to the collision integral, one estimates the 1$^{\rm st}$ order correction
\begin{equation}
\delta f_{(1)}(\epsilon_\pm) \sim  \frac{L^2/D}{\tau_E(V)} \int_{\epsilon_\pm}^{eV} \frac{d\omega}{\omega} 
\left(\frac{eV}{\omega} \right)^{(2-d)/2}
\label{F1}
\end{equation}
if $\epsilon_\pm=|\epsilon\pm eV/2|\ll eV$
and $\epsilon_\pm>\max\{E_{\rm Th},T\}$. By virtue of Pauli principle
this correction may not exceed unity, $\delta f_{(1)}\lesssim 1$,
which is true for 
$\epsilon_{\pm}\gtrsim \epsilon^*$ only, where the scale $\epsilon^*$ is given
by Eq.~(\ref{E_star}). Therefore a simple perturbation theory is valid provided
$\epsilon^*<\max\{E_{\rm Th}, T\}$. Resolving this condition we obtain that it is
the case of relatively short conductors, or small voltages, $E_{\rm Th}>1/\tau^*$.

  The time $\tau^*$ decreases with increasing the voltage and the 1$^{\rm st}$ order 
perturbation theory
finally breaks down at $E_{\rm Th}<1/\tau^*$. However, in this situation we can obtain
the result up to the factor of order of unity using the cut-off $\omega^*\simeq\epsilon^*$
in Eq.~(\ref{F_Coll}). The point is that the role of higher orders terms in the perturbation 
series is to smear the step in the distribution function $f(\epsilon)$ 
around $\epsilon=\pm eV/2 $ on the scale $\epsilon^*$, while at $\epsilon \gg \epsilon^*$
the 1$^{\rm st}$ order perturbation theory is still applicable. Therefore the overall effect is
similar to the increase of the temperature in the system. 
%$T_{\rm eff}=\epsilon^*$, in the system.

The result (\ref{S_chi}, \ref{F_Coll}, \ref{Pi}) with the cut-off $\omega^*=\max\{E_{\rm Th}, T, \epsilon^*\}$
enables to evaluate all irreducible
cumulants $\langle\!\langle n^k \rangle\!\rangle = i^k (\partial^k/\partial \chi^k )S(\chi)$ 
of a number of electrons transfered.
The expansion of $\Pi(\chi,z)$ in $\chi$ starts from $\chi^2$. Thus 
there is no correction to the current on the classical level. 
The interaction correction to the noise and to the
3$^d$ cumulant is small by the parameter $1/(E_{\rm Th}\tau_E)$ 
and it is dominated by inelastic collisions with the energy transfer $\omega\sim eV$.
On the contrary, the leading contribution to the higher order cumulants 
is due to Coulomb interaction and it is 
dominated by quasi-elastic collisions with small energy transfers 
$\omega^* \lesssim \omega \ll eV$. Up to the numerical constant the result 
is given by Eq.~(\ref{cumulants}).
The sketch of the voltage dependence for the 6$^{\rm th}$ cumulant
at different temperatures is shown in Fig.1. 
The cumulants grow with voltage in the range $E_{\rm Th}>1/\tau^*$ and
decay if $E_{\rm Th}<1/\tau^*$. The enhancement of higher order cumulants
are most strong at small temperatures $T\lesssim E_{\rm Th}$.
In this case their maximum occurs at $eV/E_{\rm Th}\sim g$ for 1D and
at $eV/E_{\rm Th}\sim g/\ln g$ for 2D.

\begin{figure}[t]
\includegraphics[width=2.5in]{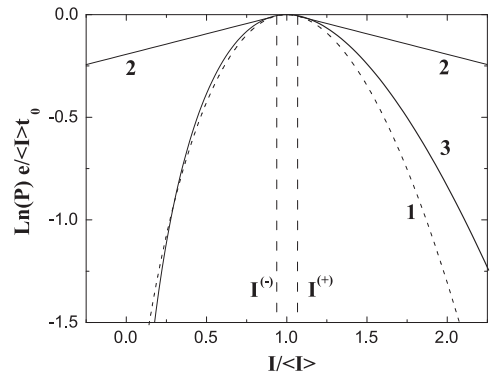}
\caption{ The log of probability to measure the big non-equilibrium current fluctuations
($eV\gg T$). Curve (1) - coherent regime, 
(2) - incoherent "cold electron" regime, $\gamma = 0.2$
(3) - "hot electron" regime.
} 
\end{figure}

  With result~(\ref{S_chi}, \ref{F_Coll}) we can also explore the current
probability distribution
$P(I)=(2\pi)^{-1}\int_{-\pi}^{\pi} \exp\{-S(\chi)+ i (I t_0/e) \chi\}$ 
in the long time limit $(I t_0/e)\gg 1$.
The action $S(\chi)$ has two branch points 
$\chi=\pm i\gamma$, where $\gamma \sim (\omega^*/eV)^{1/2}\ll 1$.
The points $\pm i \gamma$ give two threshold currents, 
$I^{\pm}=(e/t_0) \partial S/\partial\chi\bigl |_{\chi= \pm i\gamma}$, which
read $(I^{\pm} - \langle I\rangle)/\langle I\rangle = \pm \gamma /3$.

Provided $I^{-}< I < I^{+}$ the probability can be evaluated with a saddle point method.
Due to the smallness of parameter $1/E_{\rm Th}\tau_E$ we found
that $P(I)$ only slightly deviates from  the probability $P_0(I)$ of current fluctuations 
in the non-interacting limit.
For larger current fluctuations the potential $\Omega(\chi) = -S(\chi) + i (I t_0/e)\chi$
does not possess the saddle point any more and one should use
the contour $C_0$ of a zero phase, ${\rm Im}\,\Omega(\chi)\bigl|_{\chi \in C_0}=0$
for the asymptotic analysis of the integral $P(I)$. This contour 
is pinned by the branch point $\chi=\pm i\gamma$, that yields 
the {\it exponential} tails in the current probability distribution
\begin{equation}
P(I)\approx\exp\{-S(\pm\gamma) - \gamma|I| t_0/e \},\,\, I<I^{-}\, \mbox{or}\, I>I^+
\end{equation}
The results for the probability
distribution are displayed in Fig.2. The Coulomb interaction does not affect the
Gaussian fluctuations. However the tails of $P(I)$ drastically differ from
those in the absence of interaction.  
 
The FCS of this type can be understood as the statistics of the random current 
of electron-hole pairs which are excited by the low frequency fluctuations of the electromagnetic field,
produced by all other electrons in the system. 
To shed more light on this point we note that at small frequencies $\omega\ll eV$
the factor $N_\omega\cong eV/|\omega|$ in Eq.(\ref{F_Coll}) can be 
associated with non-equilibrium photon distribution.
This makes the $F_{\rm Coll}$ similar to the photocount statistics 
studied recently by Kindermann {\it et al.} in Ref.~\cite{Kindermann}. 
In the latter case each photon being transmitted through the waveguide and
absorbed by the photo-detector produces a single count.
In the given case a single absorption of a photon by the electron gas generates
the random current pulse in the circuit with a zero mean value. The generating function
of the current distribution in this pulse is given by $\Pi(\chi,z)$. Due to photon
bunching, $N_\omega \gg 1$, the pulse is strongly amplified thereby producing the
long exponential tails in the probability distribution $P(I)$.

{\it "Hot electron" regime}, $E_{\rm Th} \ll 1/\tau_E$.
In this limit the collision term in the kinetic equations dominates. 
Therefore the limiting saddle point of the action~(\ref{Eff_Action}) 
should nullify the collision integral.
To find such solution we note that the collision term in the action is
invariant under the gauge transformation 
$\widetilde{G}_\epsilon({\bf r})= e^{-\hat{K}_{\epsilon}({\bf r})} G_\epsilon({\bf r}) 
e^{ \hat{K}_{\epsilon}({\bf r}) }$. Here
$\hat{K}_{\epsilon}({\bf r}) = \frac{1}{2}\hat\tau_3[\gamma({\bf r}) + 
\beta({\bf r})(\epsilon - \phi({\bf r} ))]$ and 
$\gamma, \beta$ and $\phi$ are arbitrary functions in space. This
leads to the conservation of a current density,
and a density of the energy flow.
As well known, the physical Green function $G(\epsilon,{\bf r})$ with 
a local Fermi distribution 
$f_\epsilon({\bf r}) = [e^{(\epsilon - \phi({\bf r}))/T({\bf r})}+1]^{-1}$ nullifies
the collision term in the conventional kinetic equation. 
Its gauge transform $\widetilde{G}_\epsilon({\bf r})$ does the same for the
generalized matrix kinetic equation. 

  The unknown functions $\phi, \gamma, T$ and $\beta$ can be found from the extremum 
of the action $F_{\rm hot}$ which is obtained by substitution of 
$\widetilde{G}_\epsilon({\bf r})$
into the diffusive part of the action $F_{\rm eff}$. 
The action $F_{\rm hot}$ reads
\begin{eqnarray}
&&F_{\rm hot} = (2\pi)^{-1}g t_0 \int_0^1 dz\Bigl\{ -T(\nabla\gamma-\beta\nabla\phi)^2  
\label{F_hot}\\
&& + 
 (\nabla\gamma-\beta\nabla\phi)\nabla\phi 
-\frac{\pi^2}{3}T^3(\nabla\beta)^2 + \frac{\pi^2}{6}(\nabla T^2)\nabla\beta \Bigr\}
\nonumber
\end{eqnarray}
Here $T(z)$ and $\phi(z)$ are a local temperature and 
a chemical potential, while $\beta(z)$ and $\gamma(z)$ are their quantum counterparts.
The action~(\ref{F_hot}) implies boundary conditions:
$\phi(z)\bigl|_{z=0,1}=\pm eV/2$, $T(z)\bigl|_{z=0,1}=T$, 
$i\gamma(0)=\chi$ and $\gamma(1)=\beta(0)=\beta(1)=0$. With the
use of integrals of motion the Lagrange equations of this action 
can be reduced to two coupled second order differential equations for 
$T(z)$ and $\beta(z)$.
We are not aware of their analytical solutions under non-zero $\chi$
and solved them numerically. The results for the probability
distribution $P(I)$ are shown in Fig.~2. As in the previous section 
we have evaluated it with the use of the saddle point method. Fig.2 shows
that the probability
of positive current fluctuations, $\Delta I>0$, is enhanced in the "hot electron"
limit as compared to the coherent regime, while the probability of 
negative fluctuations, $\Delta I<0$, is affected in the lesser extent.
We also note that the action~(\ref{F_hot}) is equivalent to the actions
of Ref.~\cite{Pilgram, Mirlin}, which were derived with the use of Boltzmann-Langevin approach.
These actions transforms 
to $F_{\rm hot}$ under appropriate change of variables.

To conclude we investigated the effect of Coulomb interaction onto the FCS in the
one- and two-dimensional diffusive conductors. We have revealed the long exponential
tails in the probability of non-equilibrium big current fluctuations in short
conductors with $E_{\rm Th} \gg 1/\tau_E$, $\tau_E$ being the energy relaxation time.
These tails arise from the huge fluctuations of the current 
of electron-hole pairs which are excited by the low frequency fluctuations of \
the electromagnetic field, produced by all other electrons in the system. 
%The fluctuations are most strong for temperatures below the Thouless
%energy provided $E_{\rm Th} \sim 1/\tau_\phi$, $\tau_\phi$ being the
%voltage dependent decoherence time.
%We have also provided the quantum-mechanical derivation of the effective
%action for the FCS in the "hot electron" limit, $E_{\rm Th} \ll 1/\tau_E$.

I acknowledge the useful discussions with A.~Mirlin, Yu.~Nazarov, S.~Sharov
and A.~Zaikin. This work is a part of the CFN of the DFG and a research network of
the Landesstiftung Baden-W\"urttemberg gGmbH.

\end{document}